\documentstyle[11pt]{article}

\input{epsf}
\def\Journal#1#2#3#4{{#1} {\bf #2}, #3 (#4)}
\def\NPA{{\em Nucl.\ Phys.} A}

\def\PLB{{\em Phys.\ Lett.}  B}
\def\PR{\em Phys.\ Rev.}
\def\PRL{\em Phys.\ Rev.\ Lett.}
\def\PRD{{\em Phys.\ Rev.} D}
\def\ZPC{{\em Z.\ Phys.} C}
\def\pp{\hfill\par \vspace{\baselineskip}}

\newcommand{\mycaption}[1]{\vskip 10pt \leftskip 0pt plus 1fil
   \rightskip 0pt plus -1fil \parfillskip 0pt plus 2fil
         \baselineskip=9.5pt
         \begin{quote}
        {\noindent\footnotesize #1}\par 
         \end{quote}
         \baselineskip=13pt
}

\begin{document}

\baselineskip=16pt

\mbox{ }\hfill hep-ph/9604372 
\begin{center}
{\Large\bf Dileptons from bremsstrahlung: \\
going beyond the soft photon approximation}\footnote{
To be published in {\it International Conference on Structure
of Vacuum and Elementary Matter}, Wilderness, South Africa,
March 10--16, 1996 (World Scientific).}\\
\end{center}

\baselineskip=13pt
\begin{center}
\mbox{ }\\
H.C.\ Eggers\\
\mbox{ }\\
{\it Department of Physics, University of Stellenbosch, 
7600 Stellenbosch, South Africa} \\
\mbox{ }\\
\mbox{ }\\
C.\ Gale$^1$, R.\ Tabti$^1$, and K.\ Haglin$^2$\\
\mbox{ }\\
{\it $^1$Department of Physics, McGill University, Montr\'eal QC, 
Canada H3A 2T8}\\
{\it $^2$National Superconducting Cyclotron Laboratory, Michigan State University}\\
{\it East Lansing, MI 48824-1321, USA}\\
\mbox{ }\\
\mbox{ }\\
\end{center}

\baselineskip=11pt
\begin{quote}
{\small
The traditional calculation of dilepton yields from bremsstrahlung
relies on the assumption that electromagnetic and strong processes
factorize. We argue that this assumption, embodied by the soft photon
approximation, cannot hold true for invariant mass spectra on very
general grounds. Deriving a formula for the dilepton cross section for
pion-pion scattering that does not rely on such factorization, we
formulate the problem exactly in terms of three-particle phase space
invariants. Using a simple one boson exchange model for comparison, we
find that dilepton cross sections and yields are generally
overestimated by the soft photon approximation by factors 2--8. In
extreme cases, overestimation up to a factor 40 is possible.
}
\end{quote}

\vspace*{10mm}
\baselineskip=13pt

\section{Introduction}

Interest in the use of dileptons as a probe of the hot and early phases
of heavy ion collisions is fed by the desire of finding new physics
such as the vaunted quark gluon plasma and generally probing the
behavior of nuclear matter under extreme conditions \cite{Mul96a}.
Recent data by HELIOS and NA38 \cite{HELIOS94a} and 
discrepancies found by CERES \cite{CERES95a} have provided impetus to
hopes that new physics is finally in sight.  Realising such hopes,
however, requires detailed understanding of background processes: the
contribution from each must be calculated quantitatively.
\pp

While quantitative calculations are challenging already on a technical
level, it is a much harder problem to identify untested assumptions
that enter such calculations and to quantify their effects.  We here
aim to show, by example of bremsstrahlung from pion-pion scattering,
that the  ``soft photon approximation'' (SPA) represents such an
untested assumption \cite{Egg96a}.

\section{Why the SPA must fail}

Conventional dilepton yields at very low invariant masses $M < 500$ MeV
are dominated by Dalitz decays and bremsstrahlung. Calculations of such
bremsstrahlung yields rely heavily on the soft photon approximation
because it is simple to use.  This simplicity is achieved mainly
through the assumption that the electromagnetic and strong processes
factorize.
\pp

In order for this assumption to be valid, two conditions \cite{Low58a}
must be met:  First, the photon energy $q_0$ must be much smaller than
the energy $E_i$ of any one of the hadrons participating in the
scattering, $q_0 / E_i \ll 1$; secondly, the hadronic and
electromagnetic scales must be sufficiently different to permit
separate treatment. This translates into the condition $q_0 \ll m_Y
|\mbox{\boldmath $p_i$}|/E_i$, where $m_Y$ is the mass of the exchange
boson.  For the case where two hadrons of equal mass $m$ collide, these
equations read in their cms
\begin{eqnarray}
\label{iad}
q_0 &\ll& \sqrt{s}/2 \,, \\
\label{iae}
q_0 &\ll& m_Y \sqrt{1 - 4 m^2/s} \,.
\end{eqnarray} 
Implicit in these equations is, of course, a specific Lorentz frame
with respect to which the energies are measured.  In a simple
bremsstrahlung experiment, these limits are easily satisfied by
selecting only photons or dileptons of low energy in the laboratory
frame.  In the complex multiparticle systems formed in the course of
nucleus-nucleus collisions, however, there are many binary collisions, 
and their respective cms frames do not generally coincide either with
one another or with the overall nucleus-nucleus cms frame.
\pp

In such a situation, it is better to look at relativistically invariant
quantities, such as the dilepton invariant mass $M$. Looking at
invariant masses means that $q_0$ is no longer fixed but must vary over
its full kinematic range, which for our example of colliding equal-mass
hadrons is given by
\begin{equation}
\label{iah}
M \leq q_0 \leq {s - 4 m^2 + M^2 \over 2\sqrt{s}} \,.
\end{equation}
In Figure 1, we show the three functions (\ref{iad}) and (\ref{iae})
and (\ref{iah}) for the case $m = m_\pi = 140$ MeV, $m_Y = m_\sigma
\simeq 500$ MeV and dilepton invariant masses $M = 10$ and $300$ MeV.
It is immediately clear that the assumptions underlying the SPA are not
fulfilled even for small $M$: the kinematic range accessible to $q_0$
is never much smaller than the limits set by the SPA. The situation
becomes even worse for larger $M$.

\section{Cross sections: approximate and exact}

In order to quantify what effect the use of the SPA has on yields, it
is necessary to compare approximate cross sections to an exact
formulation. All such dilepton cross sections can  be 
written \cite{Lic95a} as the product of a purely leptonic prefactor
\[
\kappa \equiv (\alpha/ 3\pi) \left[ 1 + (2\mu^2/ M^2)\right]
\sqrt{1 - {4\mu^2/ M^2}}
\]
(with $\mu$ the lepton rest mass)
times the cross section for production of a virtual photon
$\gamma^*$ with mass $M$,
\begin{equation}
\label{brt}
d\sigma_{hh\ell^+\ell^-} = \kappa\;d\sigma_{hh\gamma^*} \,.
\end{equation}
Now the traditional procedure has been to factorise the virtual photon
cross section $d\sigma_{hh\gamma^*}$ into electromagnetic and strong
pieces, writing it in terms of a 
current $J^\mu$ and three-particle
phase space
\[
dR_3 \equiv \delta^4(p_a+p_b-p_1-p_2-q)
(d^3p_1/2E_1) (d^3p_2/2E_2) (d^3q/2q_0) \,,
\]
\begin{equation}
\label{brv}
d\sigma_{hh\gamma^*} = 4 \pi\alpha\,
{dM^2 \over M^2} 
(-J^\mu J_\mu) |{\cal M}_h|^2 \, {dR_3 \over (2\pi)^5 F}  \,,
\end{equation}
where $F$ is the incoming flux. It turns out that such factorization is
unnecessary. A re-derivation yields exactly, without 
factorization \cite{Egg96a},
\begin{equation}
\label{fcm}
d\sigma_{hh\gamma^*} =
{dM^2 \over M^2} 
\left[ - \sum_{mn} ({\cal M}_m)^\mu ({\cal M}_n^*)_\mu \right]
{dR_3 \over (2\pi)^5 F} \,,
\end{equation}
where $m,n$ run over all diagrams of the reaction
$\pi\pi \to \pi\pi\gamma^*$, including emission by the central blob.
\pp

\vspace*{4mm}

\centerline{ \epsfysize=80mm   \epsffile{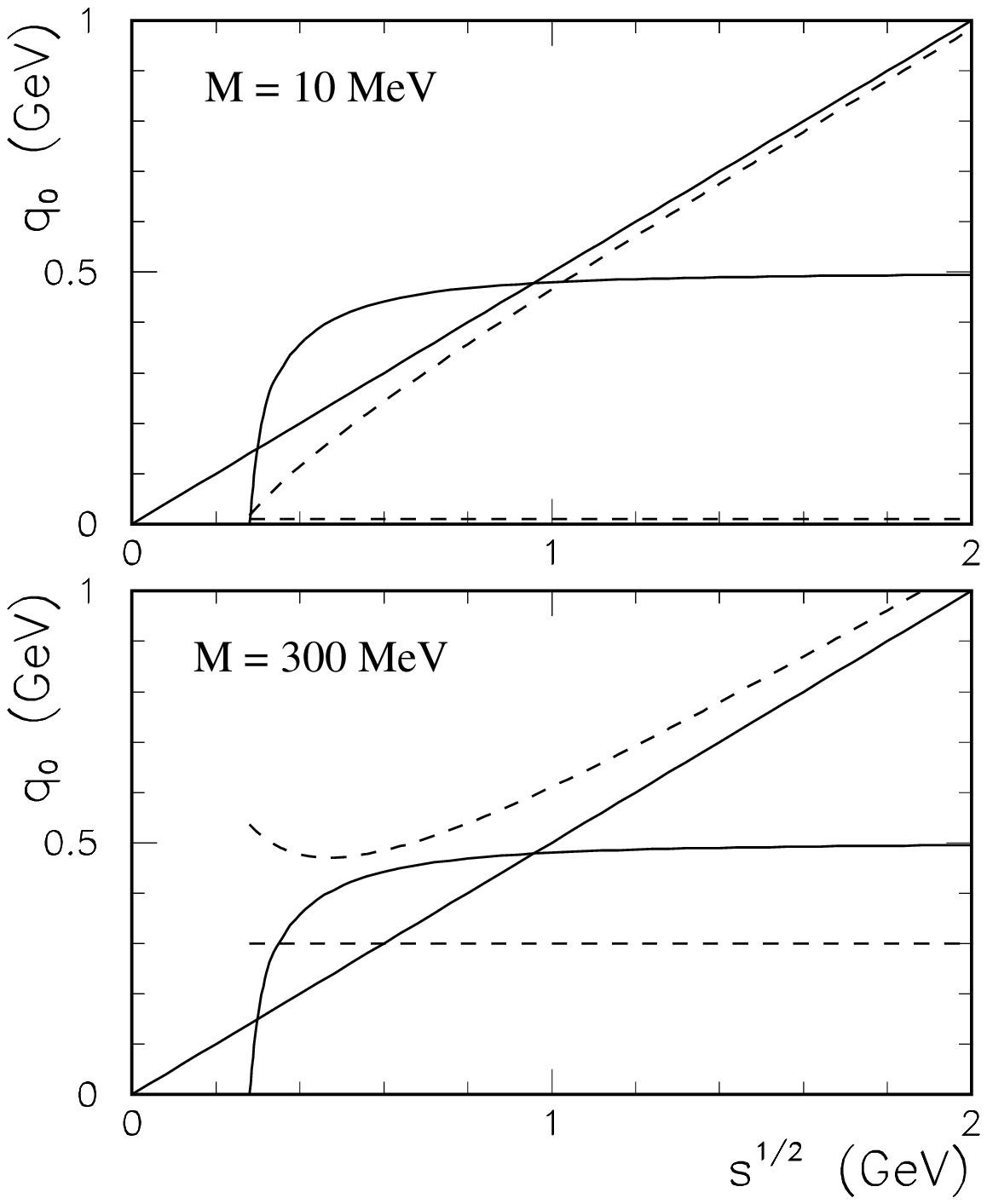}  }
{\mycaption{Figure 1:
The region between the dashed lines is the domain of integration for
$q_0$ when calculating cross sections as a function of dilepton
invariant mass.  The SPA is valid only when $q_0$ is much smaller than
the two solid lines shown. This is clearly not the case.
}}

Since people do not like 3-particle phase space,  $dR_3$
is usually simplified by neglecting $q$ in the
delta function, permitting reduction to 2-phase space
$dR_3 \simeq dR_2\; (d^3q/2q_0)$, with
$dR_2 \equiv \delta(p_a+p_b-p_1-p_2)(d^3p_1/2E_1) (d^3p_2/2E_2)$.
Eqs.\ (\ref{brt}), (\ref{brv}) then simplify to the R\"uckl 
form \cite{Ruc76a,Hag93a}
\begin{equation}
\label{rrg}
{ d\sigma_{hh\ell^+\ell^-}^{\rm Ruckl} \over dM^2 }
=  
{3\over 2} \, {\kappa \over M^2} {\alpha \over \pi} 
\int d\sigma_{hh} \, 
\int_M^A dq_0\sqrt{q_0^2 - M^2}  
\int {d\Omega_q \over 4 \pi} (-J^\mu J_\mu) \,,
\end{equation}
where $d\sigma_{hh}$ is the on-shell cross section for the purely
hadronic reaction $\pi\pi\to \pi\pi$.  The current is first
angle-averaged over the photon solid angle $d\Omega_q$ and then
integrated over the kinematic range $M \leq q_0 \leq A = [s + M^2 - 4
m^2]/2\sqrt{s}$ of the photon energy $q_0$. At this point an ad hoc
factor to correct for the factorization of $dR_3$ is usually also
included.
\pp

A better derivation by Lichard \cite{Lic95a} led to a form similar to
Eq.\ (\ref{rrg}) but without the (3/2) prefactor and with
the inclusion of $q$ and $M$ in the current.
\pp

In order to use the exact formulation (\ref{fcm}), by contrast, it is
necessary to formulate $d\sigma_{hh\gamma^*}$ in terms of {\it
three-particle phase space\/} invariants, which for the schematic
reaction $a + b \to 1 + 2 + 3$ are defined \cite{Byc73a} as
$s   = (p_a + p_b)^2$, 
$t_1 = (p_1 - p_a)^2$,
$s_1 = (p_1 + p_2)^2$,
$s_2 = (p_2 + p_3)^2$, and
$t_2 = (p_b - p_3)^2$. 
The final-state phase space integral is then given by \cite{Byc73a}
\begin{equation}
\label{fvc}
dR_3(s) =  {\pi \over 4 \lambda^{1/2}(s, m_a^2, m_b^2) }
\int {dt_1\, ds_2\, ds_1\, dt_2 \over \sqrt{B} } \,,
\end{equation}
weighted by the $6{\times}6$ Cayley determinant $B$ and where
$\lambda(x,y,z) = (x-y-z)^2 - 4yz$. One then obtains exactly
\begin{equation}
\label{xxc}
{d\sigma_{hh\ell^+ \ell^-}^{\rm exact} \over dM^2} =
{4 \pi\alpha \over (2\pi)^5 M^2}  \,
{\kappa(M^2)\pi \over 8 \lambda(s,m_a^2,m_b^2)}
\int { dt_1\,  ds_2\, ds_1\, dt_2 \over  \sqrt{B}}
\left[-\sum_{mn} ({\cal M}_m)^\mu ({\cal M}_n^*)_\mu \right],
\end{equation}
where all terms  $({\cal M}_m)^\mu ({\cal M}_n^*)_\mu$ are
written in terms of the five invariants.
\pp

As a by-product of the 3-particle phase space language, one can
define an intermediate approximation which, while still
factorising the electromagnetic part out of $d\sigma_{hh\ell^+\ell^-}$,
writes the current $J$ in terms of its invariants \cite{Egg96a}:
\begin{eqnarray}
\label{bvp}
{d\sigma_{hh\ell^+ \ell^-}\over dM^2} &\simeq&
{4 \pi\alpha \over (2\pi)^5 M^2}  \,
{\kappa\,\pi \over 8 \lambda(s,m_a^2,m_b^2)}
\int dt_1\, |{\cal M}_h(s,t_1)|^2 
\nonumber\\
&& \times
\int { ds_2\, ds_1\, dt_2 \over  \sqrt{B}}
\; \left[-J^2(s,t_1,s_2,s_1,t_2)\right]  \,.
\end{eqnarray}

\section{One boson exchange model}

Equation (\ref{xxc}) may be exact in terms of the matrix elements $\cal
M$, but it does not specify what $\cal M$ should be.  For a
quantitative comparison of the approximations (\ref{rrg}) and
(\ref{bvp}) to the exact cross section (\ref{xxc}), we must therefore
turn to a simple microscopic model of pion-pion
interactions \cite{Egg96a}.  We use a simple one boson exchange model
with the $\sigma$, $\mbox{\boldmath $\rho$}$ and $f(1270)$ mesons
included. The hadronic part of the lagrangian
\begin{equation}
\label{rtb}
{\cal L} = 
  g_\sigma\, \sigma 
   \partial_\mu \mbox{\boldmath ${\pi {\cdot}} $}
   \partial^\mu \mbox{\boldmath ${\pi}$}
+ g_\rho      \,\mbox{\boldmath ${\rho}^\mu $}
                \mbox{\boldmath ${\cdot \pi {\times}} $}
   \partial_\mu \mbox{\boldmath ${\pi}$}
+ g_f \, f_{\mu\nu} 
   \partial^\mu \mbox{\boldmath ${\pi {\cdot}}$} 
   \partial^\nu \mbox{\boldmath ${\pi}$}
\end{equation}
is fitted to elastic $\pi^+\pi^-$ collision data to fix the
constants, then supplemented through minimal substitution by
the corresponding electromagnetic interaction lagrangian.
We work at tree level only. Monopole strong form factors are
included for $t$ and $u$-channels; no electromagnetic form factors
are needed at present. Virtual photon emission is implemented
gauge-invariantly for pion and $\mbox{\boldmath $\rho$}$
emission as well as the $\pi\pi\sigma$, $\pi\pi f$ and
$\pi\pi\mbox{\boldmath $\rho$}$ vertices.
\pp

This model is probably far from a perfect description of the
pion-pion interaction, so that results obtained below can serve
merely as a good pointer towards answering the question: How different
are the results when we use the SPA or the exact cross section?

\section{Numerical results}

To quantify the differences between the approximations and the exact
formulation, we have studied the five distinct pion-pion reactions.
Writing $({+}{-}) \to ({+}{-})$ as shorthand for the reaction
$(\pi^+\pi^- \to \pi^+\pi^- \ell^+ \ell^-)$ and so on, we have
calculated within our OBE model cross sections for
$({+}{-}) \to ({+}{-})$,
$({+}{+}) \to ({+}{+})$,
$({+}{-}) \to ({0}{0})$,
$({0}{0}) \to ({+}{-})$ and
$({+}{0}) \to ({+}{0})$.
Numerical results were checked by performing  a number of consistency
checks, including testing for gauge invariance in the $\sigma$, 
$\mbox{\boldmath $\rho$}$ and $f$ sectors separately.
\pp

Figures 2--4 show the cross sections $d\sigma_{hh\ell^+\ell^-}$
for dielectron production from the five distinct reactions as
functions of
$\sqrt{s}$, for $M = 10$ MeV and 300 MeV respectively.  Final-state
symmetrization factors were included where appropriate. Initial-state
symmetrization was also included for $({+}{+}) \to ({+}{+})$ and
$({0}{0}) \to ({+}{-})$ in order to facilitate  use within a thermal
pion gas environment.
Because they are identical in structure
to their charge-conjugate versions, cross sections for the reactions
$({+}{0}) \to ({+}{0})$ and $({+}{+}) \to ({+}{+})$ were doubled.
All cross sections were computed using the
same OBE model and parameter values.
\pp

We see that there is a complete hodgepodge of curves, 
with a few underestimating the ``exact'' cross sections (solid lines)
but most overestimating the exact curves by factors  1--5.
The largest discrepancy between approximations and the exact result
occur for the reaction $({+}{+}) \to ({+}{+})$ (Figure 4):
for the R\"uckl approximation, factors 3 (for 10 MeV) to
40 (for 300 MeV) arise, while the Lichard approximation yields
corresponding overestimation factors of 1.9 and 14--20.
\pp

In Figure 5, we attempt to cast some light on the physical
origin of the discrepancies observed. Plotted are the 
cross sections for reactions $({+}{+})
\to ({+}{+})$ and $({+}{-}) \to ({0}{0})$. 
The upper lines
correspond, as before, to the R\"uckl, Lichard and 3-phase space
current approximations,
while the solid line again represents the exact result.  The lowest
dash-dotted line, on the other hand, represents the exact result but
excluding all internal \footnote{
``Internal'' diagrams are those for which the photon is emitted at
a hadronic vertex or by the exchange boson.}
diagrams and their cross terms with external ones. The difference
between this lower line and the exact result (solid line) therefore
represents the contribution of the internal diagrams; while the
difference between the lower dash-dotted line and the upper lines
(approximations) represents the change in cross section due to
inclusion/exclusion of $q$ in the {\it external}-emission diagrams.  We
see that the contribution of internal emission is not all that large,
albeit nonnegligible. By far the most important effect on $d\sigma/dM$
is the neglect of the dependence of ${\cal M}$ on the photon momentum
$q$.  We believe that this neglect is at the heart of the considerable
differences between approximations and exact results.
\pp

Finally, Figure 6 shows the dilepton production rates per unit spacetime
summed over all seven reactions, calculated using the Boltzmann 
formula \cite{Kaj86a}
\begin{equation}
\label{rtk}
{dN^{\rm Boltz}_{\ell^+ \ell^-} \over d^4x\, dM^2}
= {1\over 32 \pi^4} \int ds \, \lambda(s,m^2,m^2) 
         {K_1(\sqrt{s}/T) \over (\sqrt{s}/T)}
  {d\sigma_{hh\ell^+ \ell^-} \over dM^2 } 
\end{equation}
for temperatures $T = 100$ and 200 MeV respectively.  Right-hand panels
show the corresponding ratios, obtained by dividing a given approximate
by the ``exact'' result.  Again, the R\"uckl approximation is the
worst, as expected; overestimation factors range from 2 to 4 for $T =
100$ MeV, and 2--8 for $T = 200$ MeV.  The Lichard and 3-phase space
current approximations overestimate by factors 1.4--4, depending on
temperature and $M$.
\pp

Note that none of the approximations approaches the exact result for
small values of $M$: even for the smallest value shown ($M = 10$ MeV),
the discrepancy is still above 40\% for the Lichard and current
approximations and larger than a factor 2 for the R\"uckl
approximation.
\pp


\includegraphics{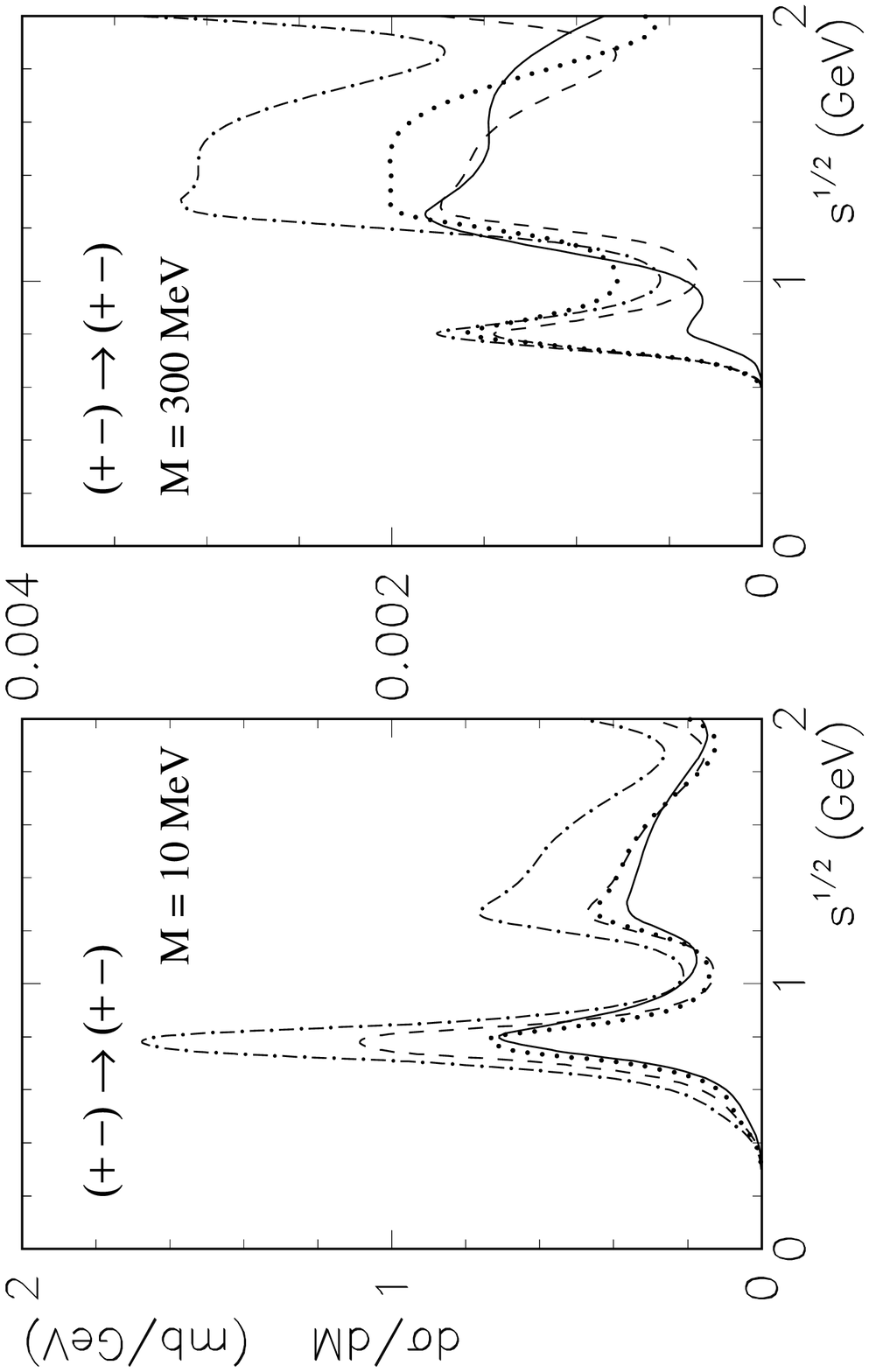}
\vspace*{75mm}
\includegraphics{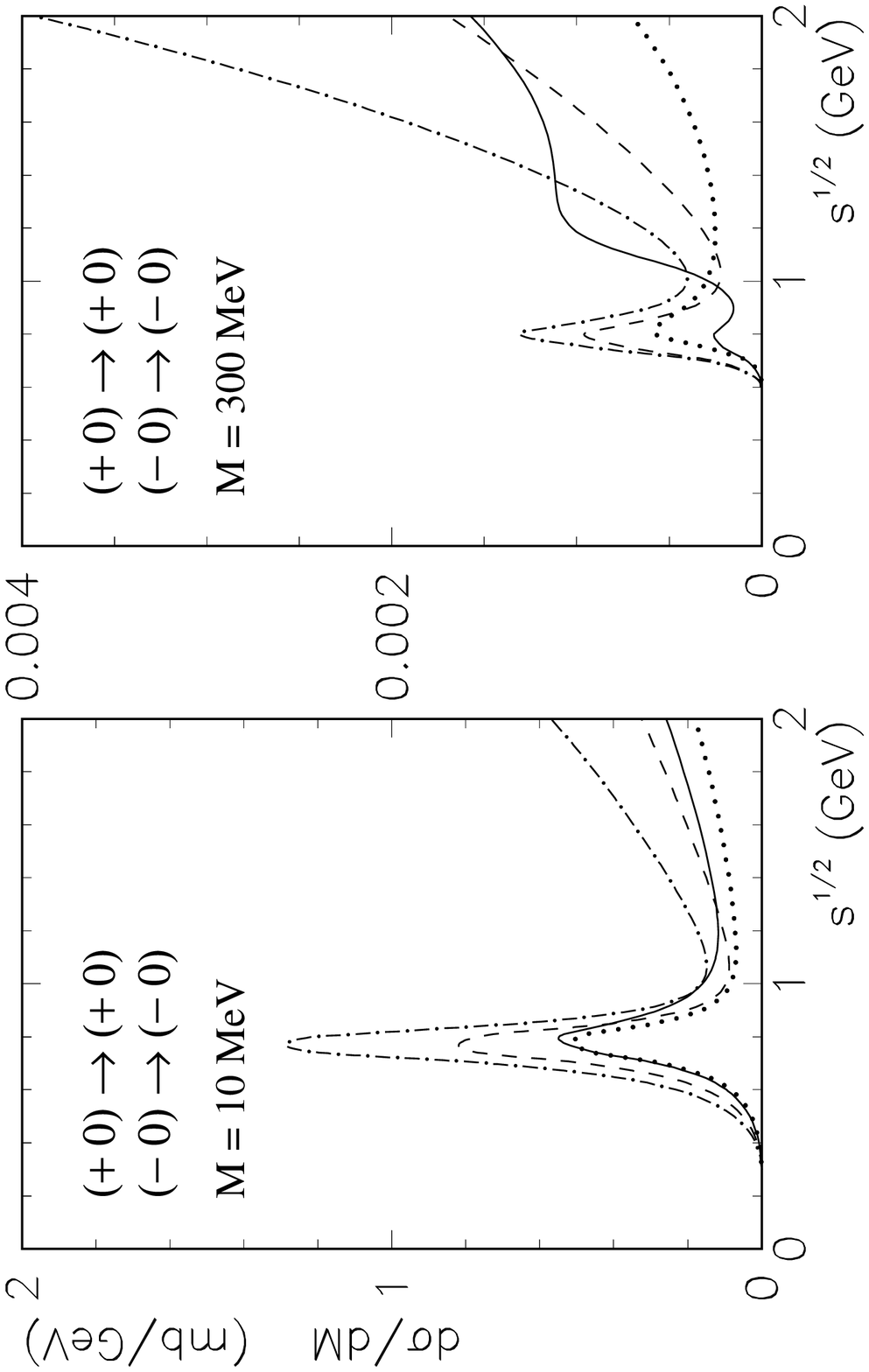}

\vspace*{78mm}
{\mycaption{Figure 2:
Cross sections for
$\pi^+\pi^- \to \pi^+\pi^- e^+ e^-$ and
$\pi^+\pi^0 \to \pi^+\pi^0 e^+ e^-$
for fixed invariant masses $M = 10$ and 300 MeV.
Solid line: exact OBE calculation Eq.\ (\ref{xxc}). 
Dash-dotted line: R\"uckl approximation (\ref{rrg}). 
Dashed line: Lichard approximation.
Dotted line: 3-phase space current (\ref{bvp}). 
}}

\newpage
\mbox{ }

\includegraphics{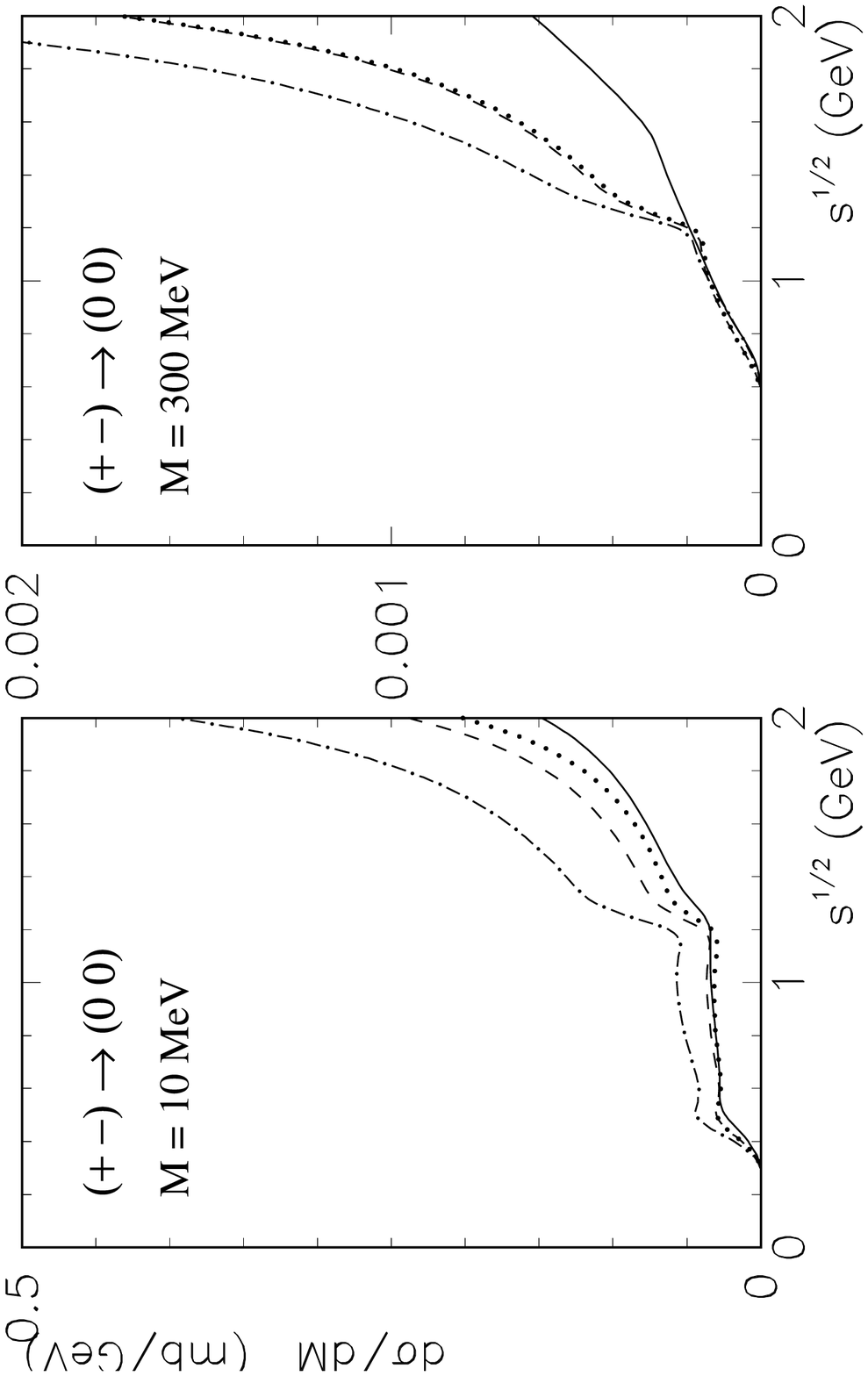}
\vspace*{75mm}
\includegraphics{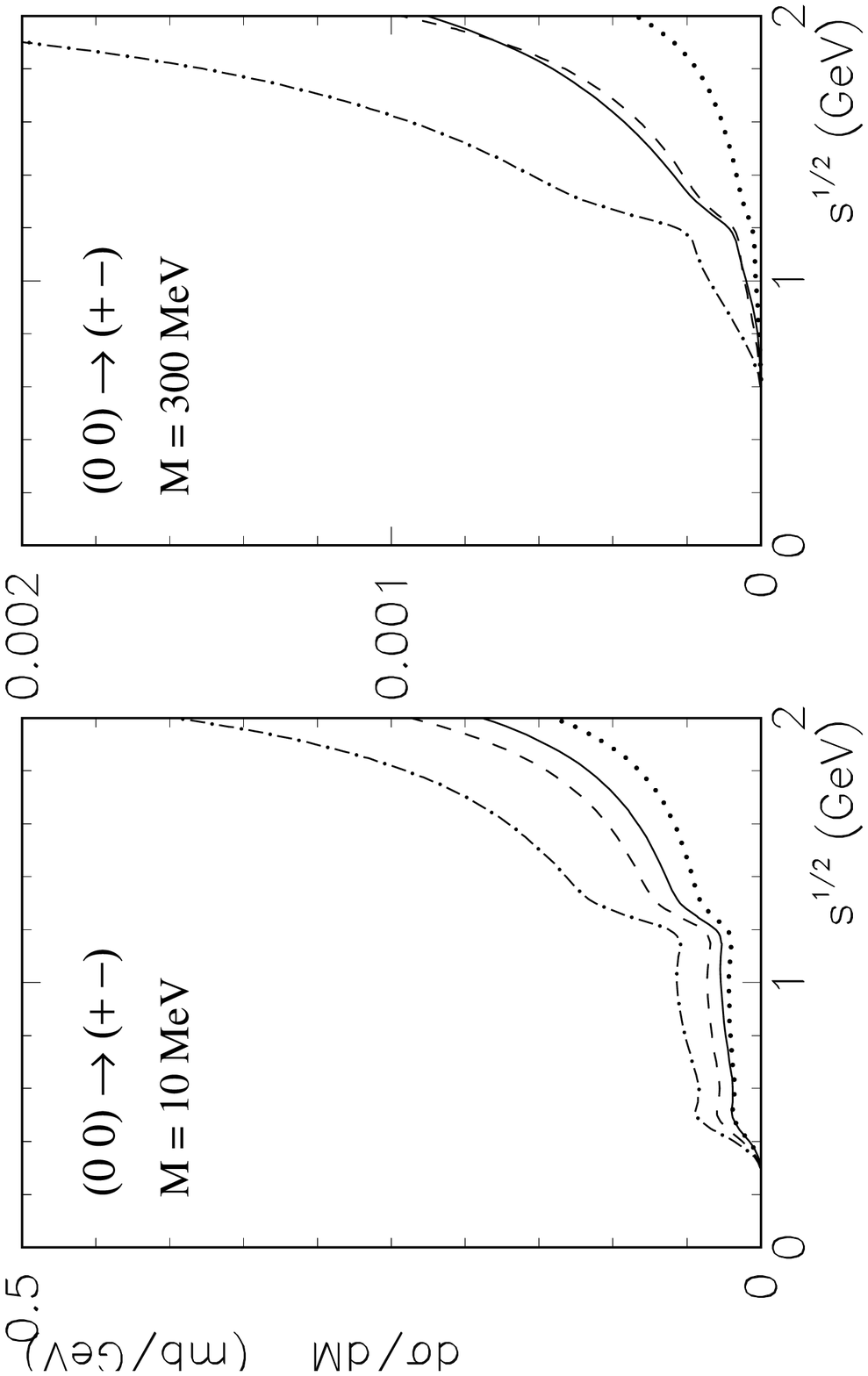}

\vspace*{78mm}
{\mycaption{Figure 3:
Same as Figure 2, for the reactions
$\pi^+\pi^- \to \pi^0\pi^0 e^+ e^-$ and
$\pi^0\pi^0 \to \pi^+\pi^- e^+ e^-$.
}}

\includegraphics{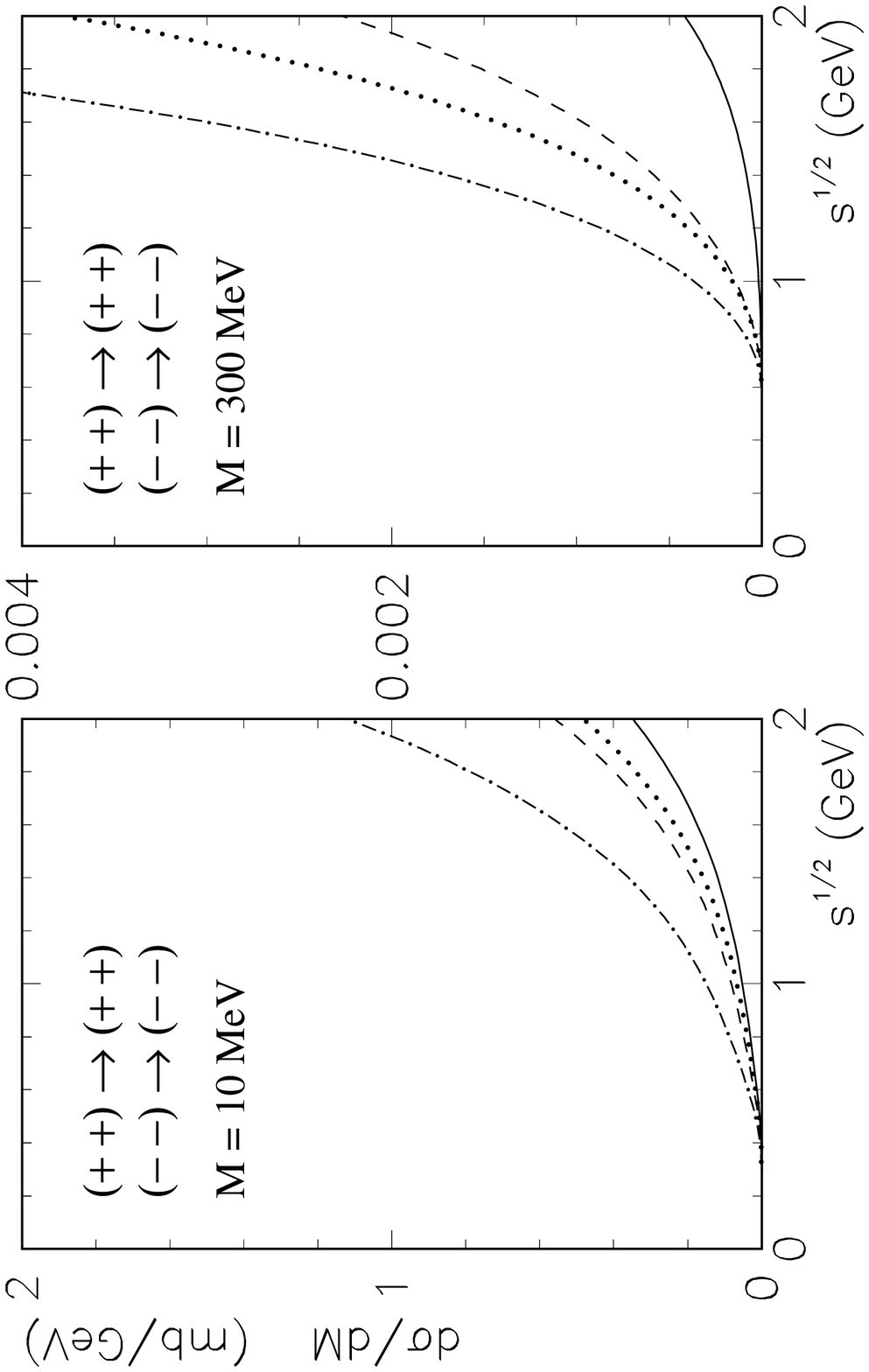}
\vspace*{64mm}
{\mycaption{Figure 4:
Same as Figure 2, for the reaction
$\pi^+\pi^+ \to \pi^+\pi^+ e^+ e^-$.
}}

\vspace*{17mm}

\includegraphics{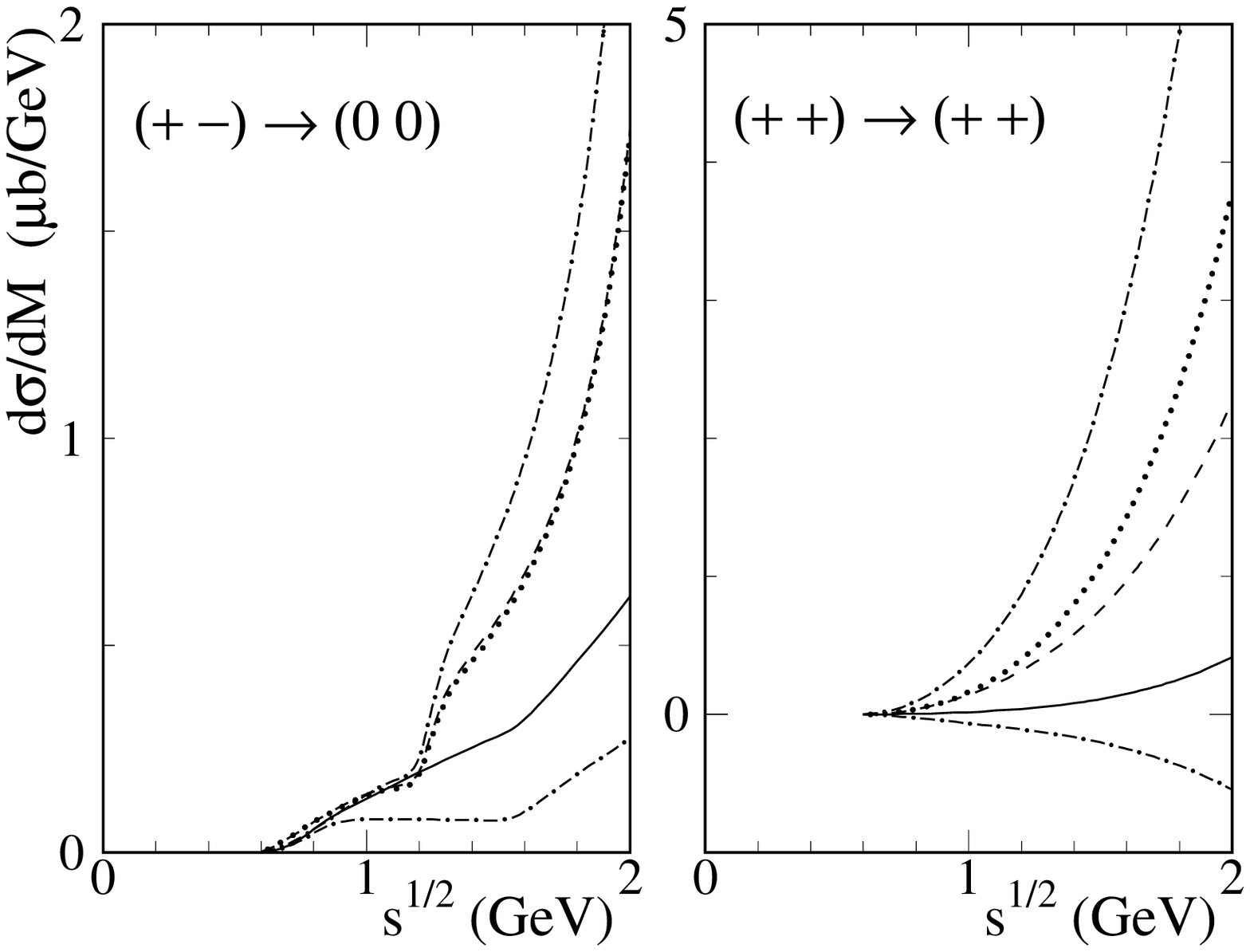}
\vspace*{68mm}
{\mycaption{Figure 5:
Contribution of external-emission vs.\ external-plus-internal
diagrams for the reactions $({+}{-}) \to ({0}{0})$
and $({+}{+}) \to ({+}{+})$ for $M = 300$ MeV.
Lines are as in Figs.\ 3 and 4. 
The new dash-dotted line below the (solid line) exact calculation
represents contributions arising solely from emission of
$\gamma^*$ by an external pion line, but taking $q$ into account
in ${\cal M}$, in contrast to the approximations (upper lines)
which neglected $q$.
}}

\begin{center}
\begin{tabular}{p{174pt}p{174pt}}
\parbox{174pt}{    
   \includegraphics{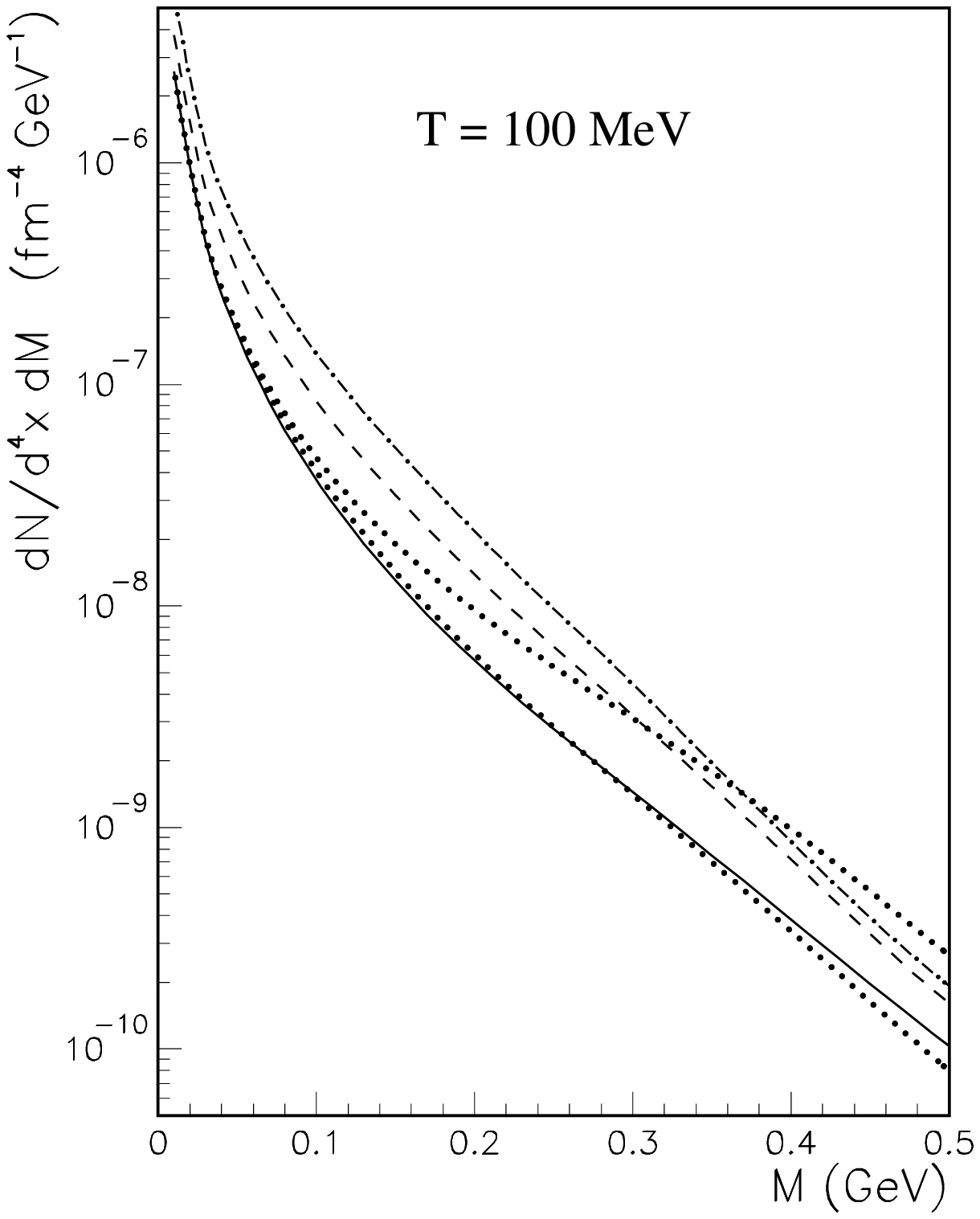}   
}
&
\parbox{174pt}{    
   \includegraphics{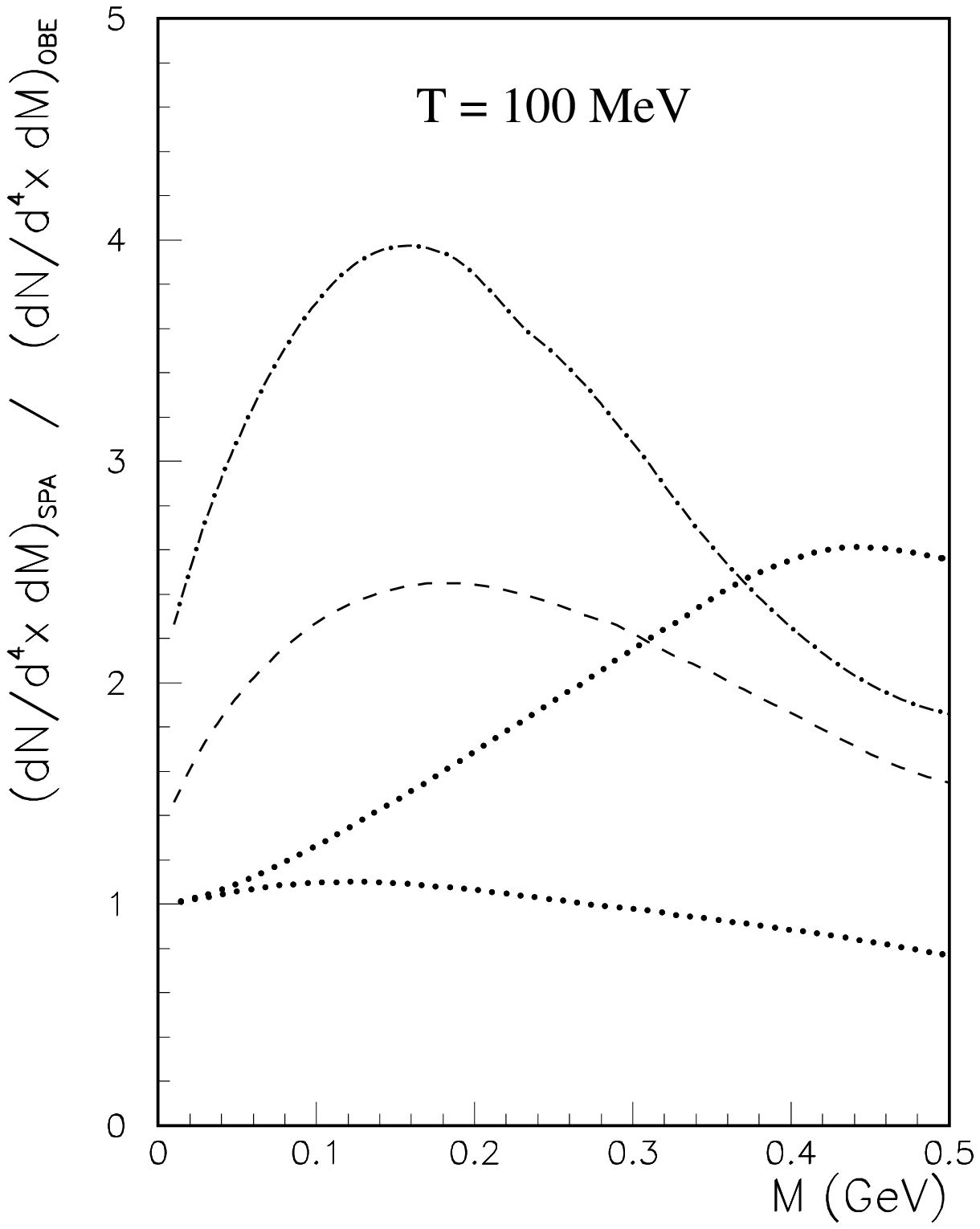}   
}
\\
\end{tabular}

\vspace*{76mm}

\begin{tabular}{p{174pt}p{174pt}}
\parbox{174pt}{    
   \includegraphics{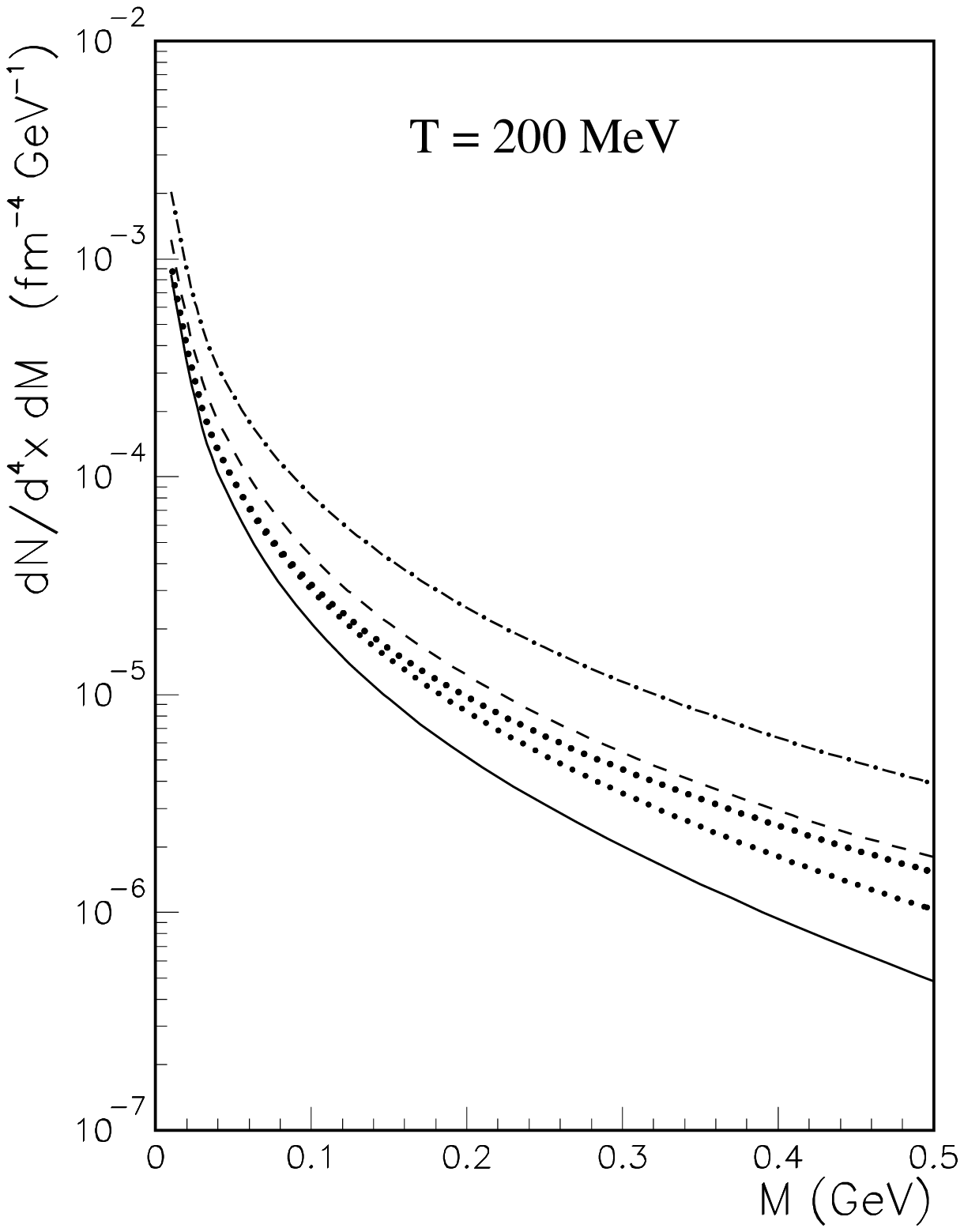}   
}
&
\parbox{174pt}{    
   \includegraphics{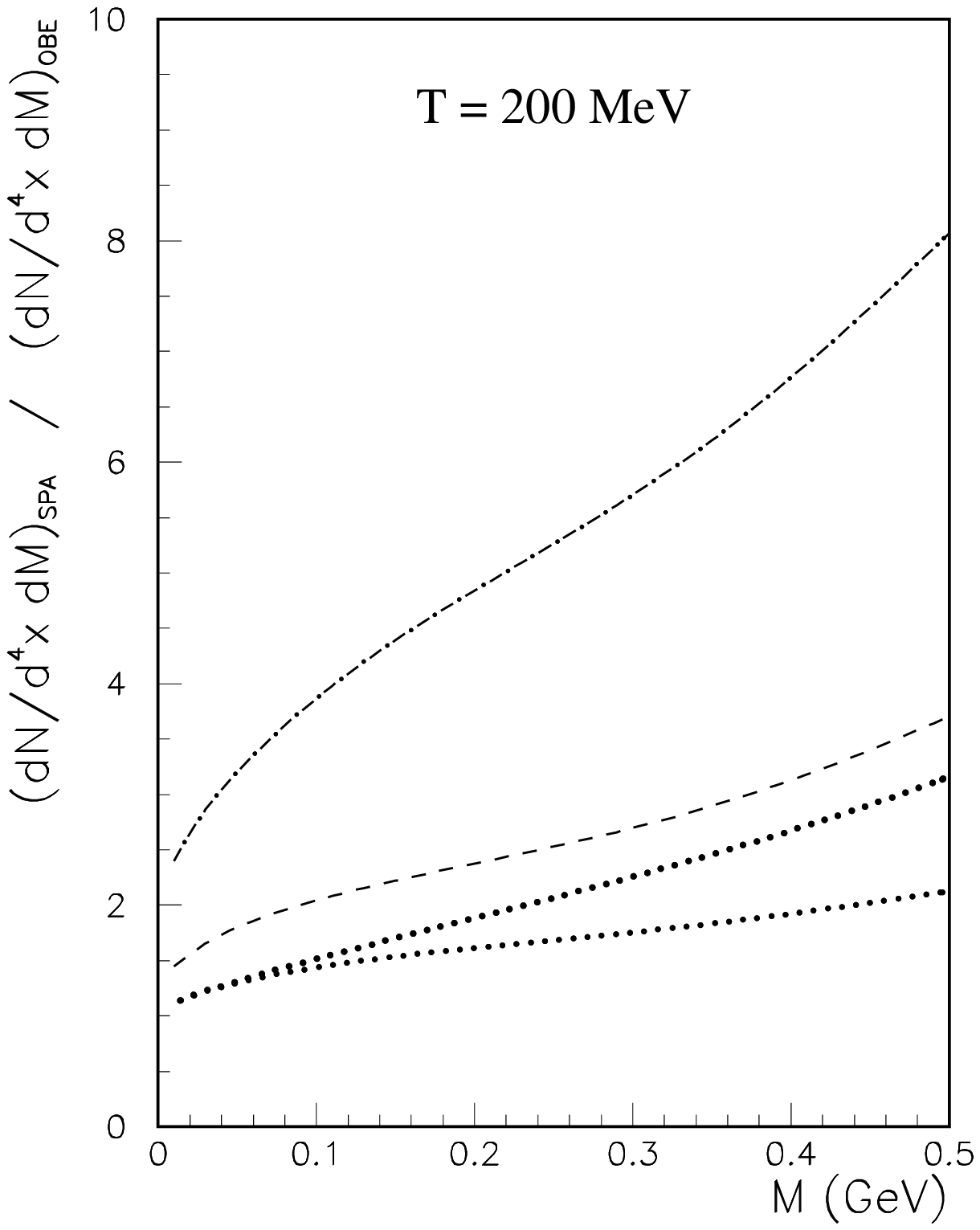}   
}
\\
\end{tabular}

\end{center}

\vspace*{68mm}

{\mycaption{Figure 6:
Left: Total bremsstrahlung yields of $e^+e^-$ for all seven
pion-pion reactions, for a Boltzmann gas with temperatures
$T = 100$ MeV and $T = 200$ MeV. 
Right: Ratios of SPA approximation calculations divided by exact OBE rate.
Solid line: exact OBE calculation, 
Dash-dotted line: R\"uckl approximation,
Dashed line: Lichard approximation,
Dotted line: 3-phase space current, including (upper) and excluding (lower) 
the mass of the virtual $\gamma^*$ in the current. 
}}

\newpage

We hence believe that the SPA is flawed when used in a heavy ion
context except in special situations. The suppression of the ``exact''
rates compared to  traditional calculations shown here would imply
that bremsstrahlung dielectrons cannot make up for the 
discrepancies between measured dielectron rates and the cocktail
of reactions used for comparison. Greater attention will have to
be paid to the $\pi\pi$ annihilation, $\eta$ Dalitz and other 
channels contributing at low $M$. The suppression we find would
also have a bearing on calculations of the Landau-Pomeranchuk 
effect \cite{Cle93b}.

\section*{Acknowledgments}
This work was supported in part by the Austrian Fonds zur F\"orderung
der wissenschaftlichen Forschung (FWF), the Natural Sciences and
Engineering Research Council of Canada, the Qu\'ebec FCAR fund, a NATO
Collaborative Research Grant, and the National Science Foundation.


\end{document}